\documentclass[pre,12pt,showpacs]{revtex4}
\usepackage{epsfig}
\begin{document}
\title{Singularity Structure in Veneziano's Model}
\author{D. Podolsky$^{\dag , \ddag}$}
\email{podolsky@cita.utoronto.ca}
\affiliation{${}^{\dag}$ CITA, University of Toronto, Toronto, ON, Canada, M5S 3H8}
\affiliation{${}^{\ddag}$ Landau Institute for Theoretical Physics, RAS, Kosygina, 2,
 Moscow, Russia, 119334}
 
\date{\today} 

\begin{abstract}
We consider the structure of the cosmological singularity in Veneziano's 
inflationary model. The problem of choosing initial data in the model is shown 
to be unsolved --- the spacetime in the asymptotically flat limit can be filled 
with an arbitrary number of gravitational and scalar field quanta. As a 
result, the universe acquires a domain structure near the singularity, with an 
anisotropic expansion of its own being realized in each domain.
\end{abstract}

\maketitle

%%%%%%%%%%%%%%%%%%%%%%%%%%%%%%%%%%%%%%%%%%%%%%%%%%%%%%%%%
\section{Introduction}
%%%%%%%%%%%%%%%%%%%%%%%%%%%%%%%%%%%%%%%%%%%%%%%%%%%%%%%%%

An effective inflationary model is known to be very
difficult to construct in the low-energy approximation
of the string theory \cite{Campbell}. The effective potentials of the
scalar field responsible for the inflationary pattern of
cosmological dynamics that arise in attempting to solve
the problem do not ensure the satisfaction of the slow-roll
condition  $|\dot{H}|\ll H^2$. Thus, the Friedman decelerating
expansion is typical of this effective gravitational
theory even in the presence of a scalar field.
Actually, this implies that the problem cannot be
solved by a brute-force method: new ideas based on the
nontrivial low-energy spectrum of the string theory or
the non-perturbative effects arising in this theory should
be invoked to obtain an inflationary scenario in terms of
the string theory. One might probably expect the field
dynamics at the inflationary phase to be also nontrivial.
The appearance of Veneziano's paper \cite{Ven1} may be
considered to be the birth time of one of these nontrivial
scenarios. The basic idea of this work is as follows.
Let us consider the low-energy effective action of an
arbitrary theory of closed superstrings \cite{GSW}, with our
analysis being restricted to the field problem with zero
vacuum averages of the fermion fields, the  $R-R$ -- sector fields
and the antisymmetric field $B_{\mu \nu}$. The sector left
after this projection has the same structure for any
superstring theory~\footnote{Recall that there are four distinct theories
of closed superstrings in ten dimensions: $IIA$, $IIB$, heterotic $SO(32)$
and heterotic $E_8$.}, and the corresponding effective
action (in dimensionless units) is
\begin{equation}
S=-\int \sqrt{-g}d^{d+1}x e^{-\phi}
\left( R+g^{\mu \nu}\partial_\mu \phi \partial_\nu \phi \right)
\label{Sven}
\end{equation}
Here,
$d$ is the dimension of space. We will seek a spatially
homogeneous classical solution to the equations of motion derived by varying 
this action. To this end, as
usual, we should set
\begin{equation}
g_{\mu \nu} = (1, -a^2 (t) \delta_{ij}),
\label{ganzv}
\end{equation}
\begin{equation}
\phi = \phi (t)
\label{fanzv}
\end{equation}
Substituting (\ref{ganzv}) and (\ref{fanzv}) into (\ref{Sven}), 
we easily find that
\begin{equation}
S = -\int d^{d+1}x a^d e^{-\phi} \left( \dot{\phi}^2 - 2dH\dot{\phi} + 
d (d-1)H^2  \right)
\label{Svennew}
\end{equation}
As can be easily seen, this action is invariant relative to
the field transformations
\begin{equation}
a(t) \to \frac{1}{a(t)},
\label{aduality}
\end{equation}
\begin{equation}
\phi \to \phi - 2d {\rm  ln } (a),
\label{phiduality}
\end{equation}
\begin{equation}
t\to -t
\label{tduality}
\end{equation}
Therefore, the equations of motion following from (\ref{Svennew})
have a classical solution that describes the accelerating
expansion corresponding to inflation at $-\infty <t< -0$
and
the Friedman decelerating expansion at  $+0 <t< +\infty$:
\begin{equation}
a_{+} (t) = (t/t_0)^{1/\sqrt{d}}, \phi = \left( \sqrt{d} - 1\right) 
{\rm ln} \left( \frac{t}{t_0}\right), t>0;
\label{a+}
\end{equation}
\begin{equation}
a_{-} (-t) = (-t/t_0)^{-1/\sqrt{d}}, \phi = -\left( \sqrt{d} + 1\right) 
{\rm ln} \left( -\frac{t}{t_0}\right), t<0
\label{a-}
\end{equation}
The physical meaning of this solution is as follows. The
universe is initially an asymptotically flat world in the
sense that the Riemann tensor components tend to zero
as $t\to -\infty$. In addition, this world is absolutely cold ---
it contains no clustered matter. Starting from this maximally
symmetric state, the universe undergoes superinflation
(the prefix ``super'' implies that $\dot{H}>0$) at  $-\infty < t < -0$
and gives way to decelerating Friedman
expansion at $t > 0$ which corresponds to the transition
from the superinflationary branch to the dual one. Significantly, as
 $t\to -0$, the spacetime curvature
tends to infinity. Therefore, sooner or later, we will go
outside the validity range for the low-energy approximation
of the string theory and action (\ref{Sven}) in our
solution. The invariance of action (\ref{Svennew}) with respect to 
transformations (\ref{aduality}) -- (\ref{tduality}) was called SF 
duality~\footnote{SF stands for the scale factor.}. 
If it holds not only
for the low-energy approximation of the string theory
but also for the total non-perturbative Green--Schwartz
sigma model, then a solution of type (\ref{a+}) and (\ref{a-}) actually 
corresponds to the saddle point in the complete field--theoretic problem;
i.e., it describes the cosmological dynamics in this
theory. However, since the SF duality itself is an essentially
non-perturbative effect. That is why we cannot ascertain
whether it exists in the non-perturbative string theory restricting ourselves
by analysis of low-energy dynamics. Gasperini and Veneziano
\cite{Ven2} argued that the non-perturbative SF duality
does takes place.
The most interesting and critical (from the viewpoint
of the scenario) point on the time axis is
$t = 0$ at
which the cosmological singularity is 
reached~\footnote{Possibly, the singularity  is smoothed out in the
non-perturbative string theory. The validity of this assumption
is closely related to the possibility of solving ``graceful exit''
problem --- the problem of passing through the singularity from
superinflation to decelerating expansion\cite{Gasperini} (see also 
\cite{Ven2}).}.
Our objective is to ascertain the pattern of field
dynamics near the singularity and understand whether an accounting
for the fluctuations of the classical trajectory  (\ref{a+}), (\ref{a-})
leads to a general softening of the singularity in
the low-energy approximation.

%%%%%%%%%%%%%%%%%%%%%%%%%%%%%%%%%%%%%%%%%%%%%%%%%%%%%%%%%%%%%%%%%%%%%%%%
\section{An anisotropic solution in Veneziano's model}
%%%%%%%%%%%%%%%%%%%%%%%%%%%%%%%%%%%%%%%%%%%%%%%%%%%%%%%%%%%%%%%%%%%%%%%%

The equations of motion that follow from the four-dimensional
theory with the action
\begin{equation}
S_{(S)}=-\frac{1}{\lambda_s^2}\int \sqrt{-g}d^4x e^{-\phi}
\left(R+(\partial \phi)^2 \right),
\label{P15}
\end{equation}
(here, we introduced the constant
$\lambda_s$ to reduce the action
to dimensionless form) are
\begin{equation}
-D_{\mu}D_{\nu}\phi+\frac{1}{2}g_{\mu \nu} (D \phi)^2
=R_{\mu
\nu}-\frac{1}{2}g_{\mu \nu}R ,
\label{P16_1}
\end{equation}
\begin{equation}
D_{\mu} D^{\mu} \phi=(\partial \phi)^2 .
\label{P16_2}
\end{equation}
Analysis of these equations can be greatly simplified if
we note that the theory described by action (\ref{P15}) is conformally
equivalent to the general theory of relativity
with the scalar field
\begin{equation}
S_{(E)}=-\frac{1}{\lambda_s^2}\int \sqrt{-g}d^4x
\left( R-\frac{1}{2}(\partial \phi)^2 \right),
\label{P17}
\end{equation}
Indeed, if the metric $g_{\mu \nu (E)}$ is the solution of the Einstein
equations, then  $g_{\mu \nu (S)}=g_{\mu \nu (E)} e^{\phi}$ is the solution
of Eqs. (\ref{P16_1}) and (\ref{P16_2}). The reverse is also true (below,
we call $g_{\mu \nu (E)}$ and $g_{\mu \nu (S)}$ the metrics in the Einstein 
and string frames, respectively).

Let us try to go beyond the scope of the homogeneous
problem and to construct an exact solution that
corresponds to a classical strong gravitational wave
propagating against a homogeneous background and
strong inhomogeneous scalar field perturbations in the
Einstein frame. Since, in general, this solution, clearly,
cannot be found, we restrict our analysis to the quasi-two-
dimensional problem. Let all metric components
and the scalar field depend on the coordinates $t$ and $x$
alone. This problem can be solved completely (see,
e.g., \cite{Kunze} , \cite{Feinstein1}), and the answer is the axisymmetric Einstein --
Rozen metric.

We seek the corresponding solution in the form
\begin{equation}
ds^2_{(E)} = e^{2A}dt^2-e^{2C}dx^2-e^{2B}(e^{\gamma}dy^2+e^{-\gamma}dz^2),
\label{P18}
\end{equation}
where $A, B, C, \gamma$ and $\phi$ are functions of $t$ and $x$ alone.
Since the Einstein equations are invariant relative to the
gauge transformations $\tilde {t}= \tilde {t}(t,x)$ and 
$\tilde {x}=\tilde {x}(t,x)$, we
may set  $g_{00}=-g_{11}, g_{01}=0$,
i.e., $A(t,x)=C(t,x)$.

The Einstein equations impose the following constraints
on the functions $A, B, \gamma$ and $\phi$:
\begin{equation}
A''-\ddot{A}-2\ddot{B}+2{\dot{A}}\dot{B}+2A'B'-2(\dot{B})^2-
\frac{(\dot{\gamma})^2}{2}={\frac{1}{2}}(\dot{\phi})^2,
\label{P19_1}
\end{equation}
\begin{equation}
-2\dot{B}'+2A'\dot{B}+2B'\dot{A}-2B'\dot{B}-\frac{\dot{\gamma}
\gamma{}'}{2}=\frac{1}{2}\phi{}'\dot{\phi},
\label{P19_2}
\end{equation}
\begin{equation}
\ddot{A}-A''-2B''+2A'B'+2\dot{A}\dot{B}-2(B')^2-
\frac{(\gamma{}')^2}{2}=\frac{1}{2}(\phi{}')^2,
\label{P19_3}
\end{equation}
\begin{equation}
\ddot{B}+2(\dot{B})^2=B''+2(B')^2
\label{P20_1}
\end{equation}
\begin{equation}
\ddot{\gamma}+2\dot{B}\dot{\gamma}={\gamma}''+2B'{\gamma}'
\label{P20_2}
\end{equation}
Finally, the equation of motion for the field $\phi$ appears as
\begin{equation}
\ddot{\phi}+2\dot{B}\dot{\phi}={\phi}''+2B'{\phi}'
\label{P21}
\end{equation}
We can easily find from (\ref{P20_1}) that
\begin{equation}
B=\frac{1}{2}{\rm ln } (f_1(\xi)+f_2(\eta)),
\label{P22}
\end{equation}
where $\xi=t-x$, $\eta=t+x$ and $f_{1,2}$ re arbitrary functions
of their argument. Let us now use the remaining gauge
invariance with respect to the transformations $\tilde{\xi}=H_1(\xi ),
\tilde{\eta}=H_2(\eta )$ and set $B=\frac{1}{2}{\rm  ln} \left(
-\frac{t}{t_i} \right)$. Equations (\ref{P20_2}) and 
(\ref{P21}) can then be easily solved:
\begin{equation}
\phi=\psi{\rm ln} (-t/t_i)+\sum_k (c_{1k} {\rm J}_0(kt)+c_{2k}{\rm
N}_0(kt))
e^{ikx}+{\rm c.c.}
\label{P23_1}
\end{equation}
\begin{equation}
\gamma=\beta{\rm ln} (-t/t_i)+\sum_k (c_{3k} {\rm J}_0(kt)+c_{4k}{\rm
N}_0(kt))
e^{ikx}+{\rm c.c.}
\label{P23_2}
\end{equation}
An expression for $A(t, x)$ can be derived by using the
equations
\begin{equation}
\dot{A}=\frac{t}{4}((\phi ')^2+(\dot{\phi})^2+(\gamma
')^2+(\dot{\gamma})^2-\frac{1}{t^2})
\label{P24_1}
\end{equation}
\begin{equation}
A'=\frac{t}{2}(\phi '\dot{\phi}+\gamma '\dot{\gamma}),
\label{P24_2}
\end{equation}
It is convenient to separate the function $A(t, x)$ into the
homogeneous and inhomogeneous parts. The former
can be easily derived from Eq. (\ref{P24_1}):
$$
A_{\rm hom}=\frac{1}{4}(\psi^2+\beta^2-1){\rm ln} (-t/t_i)+\frac{1}{4}
\sum_{k}\left( c_{1k}c_{1k}^{+}\frac{(kt)^2}{2}(J_1^2(-kt)- \right.
$$
$$
\left. J_0(-kt)J_2(-kt))+
(c_{2k}c_{1k}^{+}+c_{1k}c_{2k}^{+})\frac{(kt)^2}{4}(2J_1(-kt)N_1(-kt)-
\right.
$$
$$
\left. J_2(-kt)N_0(-kt)-J_0(-kt)N_2(-kt))+ \right.
$$
$$
\left.
c_{2k}c_{2k}^{+}\frac{(kt)^2}{2}(
N_1^2(-kt)-N_0(-kt)N_2(-kt))\right)+
$$
$$
\frac{1}{4}\sum_k (c_{1k}c_{1k}^{+}\frac{(kt)^2}{2}(J_0^2(-kt)+
J_1^2(-kt))+
$$
$$
 c_{2k}c_{2k}^{+}\frac{(kt)^2}{2}(J_0^2(-kt)-J_1^2(-kt))+
$$
$$
 (c_{1k}c_{2k}^{+}+c_{2k}c_{1k}^{+})\frac{(kt)^2}{4}(2J_0(-kt)N_0(-kt)-
J_1(-kt)N_{-1}(-kt)-
$$
\begin{equation}
 J_{-1}(-kt)N_1(-kt)) )+
(1,2)\to (3,4)
\label{P25}
\end{equation}
The inhomogeneous contribution to the function $A(t, x)$
is easier to determine from Eq. (\ref{P24_2}):
$$
A_{\rm inh}=\psi \sum_k
(c_{1k}J_0(-kt)+c_{2k}N_0(-kt))e^{ikx}+
$$
$$
\sum_{k,l}\frac{klt}{k+l}
e^{i(k+l)x}(c_{1k}J_1(-kt)+c_{2k}N_1(-kt))
(c_{1l}J_0(-lt)+c_{2l}N_0(lt)+
$$
$$
 \sum_{k,l, k\ne l}\frac {klt}{k-l}(c_{1k}J_1(-kt)+c_{2k}N_1(-kt))
(c_{1l}J_0(-lt)+c_{2l}N_0(-lt))+{\rm c.c.} +
$$
\begin{equation}
(1,2,\psi )\to (3,4,\beta )
\label{P26}
\end{equation}
Let us first consider the homogeneous limit 
($c_{\alpha k}=0$, $\forall \alpha ,k$). 
In the string frame, the spacetime metric is
$$
ds^2=\left(-\frac{t}{t_i}\right)^{\frac{1}{2}(\psi^2+\beta^2-1)+\psi}(dt^2-
dx^2)-
\left(-\frac{t}{t_i}\right)^{1+\psi+
\beta}dy^2-
$$
\begin{equation}
\left(-\frac{t}{t_i}\right)^{1+\psi-\beta}dz^2
\label{P27}
\end{equation}
The scalar curvature corresponding to this metric is
\begin{equation}
R=-\frac{\psi^2}{t^2}\left(-\frac{t}{t_i}\right)^{1/2(1-\beta^2-2\psi-\psi^2)} =
-\frac{\psi^2}{t^2_i}\left(-\frac{t_i}{t}\right)^{1/2(2+\beta^2+(\psi + 1)^2)}
\end{equation}
It tends to zero as $t\to -\infty$ for any point in parametric
space $(\psi, \beta)$. If $\beta=0$ and $\psi^2=3$, then the spacetime (the
background spacetime in the inhomogeneous problem)
is isotropic. In this case, metric (\ref{P18}) is identical to Veneziano
solution, which is asymptotically equivalent to
Minkowski flat universe for $t\to -\infty$. Below, precisely
this choice of parameters will be of particular
interest to us.

%%%%%%%%%%%%%%%%%%%%%%%%%%%%%%%%%%%%%%%%%%%%%%%%%%%%%%%%%%%%%%%%%%%
\section{The limit of asymptotically flat spacetime. Quantization}
%%%%%%%%%%%%%%%%%%%%%%%%%%%%%%%%%%%%%%%%%%%%%%%%%%%%%%%%%%%%%%%%%%%

Below, we will see that the modes in expressions (\ref{P23_1})
and  (\ref{P23_2}) can be interpreted as dilatons and gravitons
propagating against the background of curved spacetime.
The initial conditions are chosen in the following way: we
specify a sufficiently large time $t_i$ (this is the time scale which appears
in (\ref{P23_1}), (\ref{P23_2}), and(\ref{P25})) and discard all the modes 
that do not satisfy the condition $k|t_i|\gg{}1$; i.e., we neglect the
modes with a wavelength larger than the cosmological
horizon in the initial state. However, the existence of
these modes is in conflict with causality, unless the initial
state itself arose from inflation.
Below, by the limit  $t\to -\infty$, we mean all $t$ such that
$|t|\gg{}|t_i|$. In this case, the following asymptotics is possible
for all the modes without exception:
$$
J_0(-kt)\approx \sqrt{\frac{2}{\pi (-kt)}}\cos{(-kt-\pi/4)},
$$
\begin{equation}
N_0(-kt)\approx
 \sqrt{\frac{2}
{\pi(-kt)}}\sin{(-kt-\pi/4)}
\end{equation}
It thus follows that the 
``correct''~\footnote{I.e., those corresponding to the positive frequency 
solutions.} modes are
$$
\frac{1}{2}e^{i\pi/4}H_0^{(1)}(-kt)e^{-ikx}
$$ 
and 
$$
\frac{1}{2} e^{i\pi/4}H_0^{(1)+}(-kt)e^{ikx}.
$$
which correspond to the substitutions
\begin{equation}
b_{1k}=e^{i\pi /4}(c_{1k}-ic_{2k}), b_{2k}=e^{i\pi
/4}(c_{1k}^{+}-ic_{2k}^{+})
\label{P28}
\end{equation}
Let us now turn to quantization. To properly normalize the modes
$u_k=\frac{1}{2}e^{i\pi/4}H_0^{(1)}(-kt)e^{ikx}$
we calculate the commutator $[ b_k, b_{k'}^{+} ]$. Assuming that 
\begin{equation}
[\phi (t,x), \phi (t,x')]=0, [\pi (t,x), \pi (t,x')]=0,
\end{equation}
\begin{equation}
[\phi (t,x), \pi (t,x')]=-\frac{i}{L^2}\delta (x-x') ,
\label{P31}
\end{equation}
where $L$ is the infrared cut-off (linear size of the system under 
consideration) and
$$
\pi (t,x) = \frac{\delta \sqrt{-g}L}{\delta \dot{\phi}} =
-\frac{2}{\lambda_s^2}\sqrt{-g} e^{-\phi} g^{00} \dot{\phi} ,
$$ 
we can easily find that 
\begin{equation}
[ b_k, b_{k'}^{+} ]=\frac{\pi \lambda_s^2 t_i}{2L^3}\delta_{kk'}
\label{P32}
\end{equation}
and the correctly (half-quantum) normalized modes for $t\to -\infty$ are
$$
u_k =\sqrt{\frac{-\lambda^2_s t_i}{L^3 k t}}e^{-ik(x-t)}.
$$

To physically interpret these quantum modes, we
calculate  $\langle T_{00} \rangle$. In the Einstein equations, the 
$(0, 0)$--component is
\begin{equation}
4\dot{A}\dot{B}-2\ddot{B}-2\dot{B}^2=\frac{1}{2}((\dot{\phi})^2+
(\phi ')^2+(\dot{\gamma})^2+(\gamma ')^2)=T_{00}
\end{equation}
(we carried over $\gamma$'s to the right and now interpret this
part of the metric as the contribution of gravitons).
If
the paired correlators are assumed (in this case, it does
not matter whether we consider the amplitudes of the
modes $b_{\alpha k}$ as classical but randomly distributed
Gaussian variables or as operators) to be
\begin{equation}
\langle b_{\alpha k}b_{\beta l}^{+} \rangle = n_{\alpha}(k)\delta_{\alpha
\beta}\delta_{kl},
\langle b_{\alpha k}b_{\beta l} \rangle = 0
\label{P33_1}
\end{equation}
\begin{equation}
n_1 (k)=n_2(k), n_3(k)=n_4(k)
\label{P33_2}
\end{equation}
(the physical meaning of this condition is that the flux
of rightward-propagating quanta is equal to the flux of
leftward-propagating quanta), then
\begin{equation}
\langle T_{00} \rangle = \langle \hat{H} \rangle =\sum_k
\frac{\lambda_s^2 kt_i}{L^3 t}\left( n_1(k)+ n_3(k) \right)
\label{P34}
\end{equation}
Thus, we see that for $t\to -\infty$, the asymptotically flat
world is filled with gravitational and scalar field quanta,
which are naturally called gravitons and dilatons,
respectively. Their energy density tends to zero as 
$t\to -\infty$ (which is nothing  than usual redshift) and can not be neglected.
The cosmological solutions
with zero occupation numbers have a zero measure
in the functional space of the field-theoretic problem and, in this
sense, are untypical. Thus, there is an infinite arbitrariness
in choosing the initial state for Veneziano model,
and the problem of initial data is yet to be solved
(see in this context \cite{LindeV}).

%%%%%%%%%%%%%%%%%%%%%%%%%%%%%%%%%%%%%%%%%%%%%%%%%%%%%%%%%%%%%%%
\section{Spacetime structure near the singularity}
%%%%%%%%%%%%%%%%%%%%%%%%%%%%%%%%%%%%%%%%%%%%%%%%%%%%%%%%%%%%%%%

In the limit $t\to -0$, the spacetime metric in the
string frame is asymptotically equivalent to the Kasner
metric with the indices
\begin{equation}
p_1=\frac{1/2((\beta+\beta_1)^2+(\psi+\psi_1)^2-1)+\psi+\psi_1}
{1/2((\beta+\beta_1)^2+(\psi+\psi_1)^2+3)+\psi+\psi_1},
\label{P40}
\end{equation}
\begin{equation}
p_2=\frac{1+\psi+\psi_1+\beta+\beta_1}{1/2((\beta+\beta_1)^2+(\psi+\psi_1)^2+3)
+\psi+\psi_1},
\label{P41}
\end{equation}
\begin{equation}
p_3=\frac{1+\psi+\psi_1-\beta-\beta_1}
{1/2((\beta+\beta_1)^2+(\psi+\psi_1)^2+3)+\psi+\psi_1},
\label{P42}
\end{equation}
where
\begin{equation}
\psi_1=\sum_k \frac{2}{\pi}N
\left(e^{i\frac{\pi}{4}}(b_{1k}e^{ikx}+b_{2k}e^{-ikx})+{\rm c.c.}\right)
\end{equation}
\begin{equation}
\beta_1=\sum_k \frac{2}{\pi}N
\left( e^{i\frac{\pi}{4}}(b_{3k}e^{ikx}+b_{4k}e^{-ikx})+{\rm c.c.}\right),
\end{equation}
and  $N = \sqrt{\frac{\pi \lambda_s^2 t_i}{2L^3}}$ is the normalization constant for
the modes calculated in the preceding section.
The contribution to the action of dilaton $\phi$ from the
part of the classical trajectory (\ref{P23_1}) in the vicinity of the
cosmological singularity $\Delta S_{\rm sing}$ logarithmically
diverges --- it is proportional to $ {\rm ln} \left( \frac{ \Lambda}{t_0} \right)$, 
where $\Lambda$ is the
beginning of the Kasner epoch, and  $t_0\to 0$ is some
small time scale. At the same time, the remaining contribution
to the action is finite. This can be interpreted as the
destruction of quantum coherence~\cite{ascog} between the
modes  $c_{1k}$ and $c_{2k}$ ($c_{3k}$ and $c_{4k}$) for which the condition
$|kt| \ll 1$ is satisfied. Because of this destruction, the
modes with a wavelength larger than the cosmological
horizon freeze --- their amplitudes may be considered to
be classical, randomly distributed variables rather than
operators~\footnote{Strictly speaking, this can be done if the effective occupation
numbers satisfy the condition $\langle n_k \rangle \gg 1$. 
In this case, the non-commutativity of the coordinates
and momenta may be disregarded. For a more detailed
discussion, see \cite{Polarski}.}. These modes contribute to the Kasner indices.
Thus, the spacetime structure
becomes stochastic as $t\to -0$, and it makes sense to
discuss the behavior of the various correlation functions
of the Kasner indices. Below, we disregard the
time-independent contributions to the metric, because
they bear no relation to the character of the metric singularity
for  $t\to -0$.

Technically, it is more convenient not to pass to
world time but work with a metric of the form $ds^2 = t^{q_1}(dt^2-dx^2) -
t^{q_2}dy^2 - t^{q_3}dz^2$.
The quantities  $q_1, q_2$ and $q_3$ are related to the Kasner
indices by
\begin{equation}
q_1=\frac{2p_1}{1-p_1}, q_2=\frac{2p_2}{1-p_1}, q_3=\frac{2p_3}{1-p_1}
\label{P43}
\end{equation}
Since the identity $p_1^2+p_2^2+p_3^2 = 1$ takes place in the string
frame, the following relation is also valid:
\begin{equation}
q_2^2+q_3^2 = 4(q_1+1)
\label{P44}
\end{equation}
he stochastic spacetime structure can be completely
determined by calculating the distribution function
$F(\lambda, \mu) = \langle \delta(q_2-\lambda) \delta(q_3-\mu) \rangle=
\int dx{} dy{} e^{-i(\lambda x + \mu y)} \langle e^{i(q_2 x + q_3 y)} \rangle$.
Since $q_1$ can be unambiguously determined from the
known quantities $q_2$ and $q_3$, this distribution function
allows expressions for any correlators of the indices $q_i$
to be derived. After simple calculations, we obtain
\begin{equation}
F(\lambda, \mu) = \frac{\pi}{\sqrt{N_1 N_3}} {\rm exp}
\left(-\frac{(1+\psi-\frac{\lambda+\mu}{2})^2}{2N_1}
-\frac{(\beta-\frac{\lambda-\mu}{2})^2}{2N_3}
\right)
\label{P47}
\end{equation}
where $N_1=\sum_k \frac{n_{1k}N^2}{\pi^2}$ and $N_3=\sum_k
\frac{n_{3k}N^2}{\pi^2}$.

This function characterizes the spacetime ``in the
infinitely small''. Since the problem is translationally
invariant, all points in space are equal in rights: the
locally measured Kasner indices are the random variables
described by the distribution function (\ref{P47}). However,
the probability that the Kasner indices are small at
a given point and large at a infinitely close point approaches zero.
The nonlocal correlation
properties of the Kasner indices are specified in part by
the two-point correlation functions~\footnote{The remaining two-point correlation 
functions can be easily calculated
by using Eq. (\ref{P44}), which relates the Kasner indices. We do
not give the corresponding expressions here, because they are too
cumbersome.}
\begin{equation}
\langle q_2(x) q_2(x')\rangle = (1+\psi+\beta)^2 + \sum_k \frac{16 N^2
(n_{1k}+n_{3k})}{\pi^2}\cos {k(x-x')},
\label{P48}
\end{equation}
\begin{equation}
\langle q_3(x) q_3(x') \rangle = (1+\psi-\beta)^2 + \sum_k \frac{16 N^2
(n_{1k}+n_{3k})}{\pi^2}\cos {k(x-x')},
\label{P49}
\end{equation}
\begin{equation}
\langle q_2(x) q_3(x') \rangle = (1+\psi)^2 - \beta^2 + \sum_k \frac{16 N^2
(n_{1k}-n_{3k})}{\pi^2}\cos {k(x-x')}
\label{P50}
\end{equation}

We can see that the correlation is oscillatory in pattern,
which is an artifact of specifying the
initial conditions for $t\to -\infty$.

In conclusion, let us consider the transition possibility
from the superinflationary expansion of the universe
to its contraction as the cosmological singularity is
approached.

If the spacetime metric is isotropic and uniform,
then we say that the universe undergoes superinflationary
expansion when the condition 
$\dot{H} = \frac{\ddot{a}}{a} - \left( \frac{\dot{a}}{a} \right)^2 > 0$,
is satisfied,
where $a$ is the scale factor. If the latter changes with
time as $a= (-t)^p$, then the inequality is satisfied for $p < 0$.
This criterion can be generalized to the anisotropic
case in two ways (as we will see below, the results
depend only slightly on the choice of a criterion):

(i) by introducing not one but three scale factors and
requiring the satisfaction of the inequalities $p_1 < 0$,
$p_2 < 0$, and $p_3 < 0$;

(ii) by requiring the satisfaction of the inequality
$p_1 + p_2 + p_3 < 0$, which characterizes the behavior of the
comoving 4-volume element.

Since the spacetime acquires a stochastic structure
that is locally characterized by distribution (\ref{P47}) as the
singularity is approached, there is a nonzero probability
that superinflation will stop and will give way to contraction.
It is easy to see that, according to criterion (i),
the inverse (in a sense) quantity --- the probability that
superinflation continues until the fall of the universe to
a singularity~\footnote{This type of singularity is characterized
by tending the effective gravitational constant to infinity.} --- is
$$
w (\forall p_i < 0) = w (\forall q_i < 0) = \frac{1}{2\pi \sqrt{N_1 N_3}}
\int_0^{\sqrt{2}} \rho d\rho \times
$$
\begin{equation}
\int_{3\pi/4}^{5\pi/4} d\phi \exp
\left( -\frac{(1+\psi-\rho \cos {\phi})^2}{2N_1}
- \frac{(\beta-\rho \cos {\phi})^2}{2N_3}
\right)
\label{P54}
\end{equation}
In the physically interesting case with  $N_1, N_3 \to \infty$ and
$\psi , \beta \sim 1$ we have
\begin{equation}
w (\forall p_i < 0) \sim \frac{1}{8\sqrt{N_1 N_3}}
\label{P55}
\end{equation}
The probability that superinflationary expansion will
give way to contraction in all directions in the same
limit behaves as
\begin{equation}
w (\forall p_i > 0) \sim \frac{1}{\pi} \arctan {\sqrt{\frac{N_1}{N_3}}}
 - \frac{1}{8\sqrt{N_1 N_3}}
\label{P56}
\end{equation}

If, however, we proceed from criterion (ii), then the
probability of maximum superinflation duration is
$$
w (\sum p_i < 0) = \frac{1}{2\pi \sqrt{N_1 N_3}} \int_0^{\sqrt {6}} \rho
d\rho \int_0^{2\pi} d\phi \exp {\left(
-\frac{(3+\psi-\rho \cos {\phi})^2}{2N_1}
\right. }
$$
\begin{equation}
{\left.
- \frac{(\beta - \rho \sin
{\phi})^2}{2N_3}
\right)} \sim \frac{3}{\sqrt{N_1 N_3}} (N_1, N_3 \to \infty; \psi, \beta
\sim 1)
\label{P57}
\end{equation}

Since asymptotics (\ref{P54}) and (\ref{P57}) are virtually independent
of the choice of a superinflation duration criterion,
we can say that superinflationary expansion necessarily
gives way to contraction as the cosmological
singularity is approached.

\begin{figure}
\begin{center}
\epsfig{file=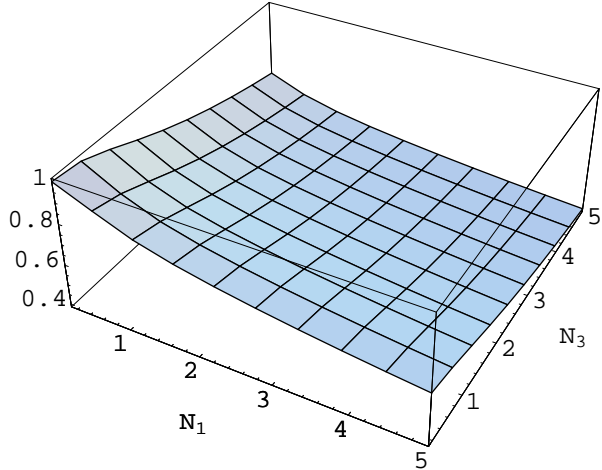,width=80mm}
\end{center}
\caption{Behavior of the probability of maximum superinflation
duration, $w(\sum p_i <0)$, as a function of $N_1$ and $N_3$ for
$\psi=-\sqrt{3}, \beta=0$ at relatively low graviton and dilaton
densities in the initial state.}
\label{Fig-ven1}
\end{figure}

\begin{figure}
\begin{center}
\epsfig{file=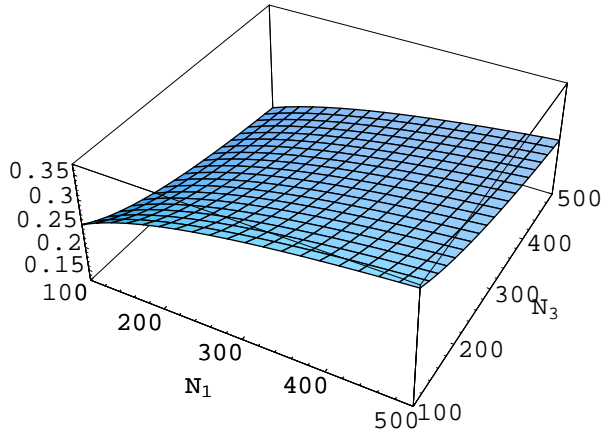,width=80mm}
\end{center}
\caption{Behavior of the transition probability from superinflationary expansion
to contraction in all directions,
$w(\forall p_i >0)$, as a function of $N_1$ and $N_3$ for $\psi=-\sqrt{3}, \beta=0$
and high energy densities of gravitons $N_1$ and dilatons
$N_3$ in the initial state.}
\label{Fig-ven2}
\end{figure}

%%%%%%%%%%%%%%%%%%%%%%%%%%%%%%%%%%%%%%%%%%%%%%%%%%%%%%%
\section{Conclusions}
%%%%%%%%%%%%%%%%%%%%%%%%%%%%%%%%%%%%%%%%%%%%%%%%%%%%%%%

We have considered the structure of the cosmological
singularity in Veneziano's model. As we showed,
the problem of uniqueness in choosing the initial conditions
in Veneziano's scenario is yet to be solved --- the
asymptotically flat world that corresponds to the initial
state in the scenario can be filled with gravitational and
scalar field quanta. In order to understand what influence
any variations in initial data have on the spacetime
structure near the singularity, we constructed an exact
solution that described the gravitational and scalar field
quanta propagating against the background of asymptotically
flat spacetime in the limit $t\to -\infty$. 

The first essentially new effect that we have found is 
the hard imprint of initial conditions on the structure of
spacetime near the singularity. 
Primordial quantum fluctuations of gravitational and matter fields  
cause the spacetime to acquire domain structure, this probabilistic 
domain structure being completely determined by the correlations of fields 
in the initial state for superinflation.
Indeed, though we deal with a strong gravitational background near the
singularity, effects of particle creation are not important
for the large-scale 
structure~\footnote{``Large scale'' means any scale larger then
$R^{-1/2}$, where $R$ is the scalar curvature of spacetime.} of spacetime 
at $t=0$.  Near the singularity 
the characteristic wavelength of created particles $\lambda \sim R^{-1/2}$ is 
much less then the typical size of domain $1/k$ --- a constant defined by 
initial conditions for the scenario. Also, back-reaction of created particles 
can be neglected until we reach the string scale~\footnote{The exact criterion 
is $\lambda_s^2 e^\phi R^{-1} < 1$, see
for details \cite{Ven2}.}, where the low-energy approximation is no longer 
valid. That is why created particles are not important, and large-scale
structure of spacetime near the singularity is completely described by initial conditions,
i.e., number of particles in the initial state for superinflation. 

Our another goal was to show once again that   
Kasner-like solution of Einstein equations found by Belinskii, Lifshitz and 
Khalatnikov \cite{BKL} has the generic character near cosmological singularity. It is well known that if
the matter sector of underlying theory contains only hydrodynamic types of 
matter, the approach toward singularity acquires chaotic features.
Generally speaking, the situation is different, if we take into account such
relativistic forms of matter as scalar field (see \cite{BK}, \cite{Dabr3}) --- the generic
solution of Einstein equations becomes to exhibit a non-oscillatory
power-law behavior. Nevertheless, it has 
been shown \cite{Damour}, that BKL chaos near singularity is resuscitated if
the underlying theory of gravity is described by bosonic sector of low-energy 
string effective action.~\footnote{See also \cite{Dabr4} for the discussion of properties of Kasner-like solutions in 
the Horava--Witten cosmology.} It should be noted that chaotic structure of spacetime
found in this note bears no relation to BKL chaos. The large-scale domain 
structure of spacetime near the singularity is completely 
determined by quantum initial conditions for the scenario. In a sense, this 
chaos is quantum, not classical one.

The physical reason for this chaotic behavior to appear is that the radius
of the cosmological horizon $R^{-1/2}$ specifies the causal
connectivity scale in the theory; accordingly, the quantum
quantities can correlate only on scales smaller than
$R^{-1/2}$. Freezing of amplitudes of initially quantum modes implies that an observer living
at the Kasner epoch will always record the same Kasner
indices, irrespective of the number of experiments that
he or she carries out. Nevertheless, before the time for the
Kasner asymptotics to become valid, we cannot predict
with confidence what Kasner indices the
observer will record --- in this sense, the Kasner indices
are random variables.

The second new effect that we have found in this note is the general softening of 
singularity due to the account of quantum initial conditions variety in the 
sense that at $t\to -0$ the spacetime dynamics tends to transition from 
superinflationary expansion to contraction. 
We believe that this fact can have a bearing on solving ``the graceful exit'' problem in 
Veneziano's string cosmology. Nevertheless, the regime in which this contraction is realized
turns to be rather complicated.

Fig. \ref{Fig-ven1} shows the behavior of the probability of
maximum superinflation duration $w(\sum p_i <0)$ as a
function of total number of dilatons and gravitons in the initial state ($N_1$ and $N_3$, respectively) 
for  $\psi=-\sqrt{3}$, $\beta=0$, which
corresponds to the absence of a seed anisotropy, i.e.,
Veneziano's universe. Interestingly, at moderately large
$N_1$ and $N_3$, an increase in the graviton energy density
with respect to the dilaton energy density generally
causes this probability to decrease. Nevertheless, at
large $N_1$ and $N_3$, the graviton and dilaton energy densities
have the same weight from the viewpoint of their
influence on the probability of maximum superinflation
duration --- an increase in the number of gravitons and
dilatons in the initial state always causes this probability
to decrease.

When considering the asymptotics of the transition
probability from superinflation to contraction in all
directions, $w(\forall p_i >0)$ at large $N_1$ and $N_2$, we arrive at a
similar picture (see Fig. \ref{Fig-ven2}): an increase in the number
of gravitons in the initial state generally causes this
probability to decrease, while an increase in the number
of dilatons causes it to increase. This implies that the
statistical weight of the states describing anisotropic
expansion (when there is expansion in two of the three
directions in space and contraction in the third direction,
etc.) becomes large; the higher the graviton energy
density in the initial state, the larger this weight.

Finally, let us discuss the degree of generality of our results. To determine the structure
of spacetime near the singularity, we actually used the simplified model describing quasi-two-dimensional
dynamics of the four-dimensional theory.  From the physical point of view this simplification means that initial conditions
at asymptotically flat infinity $t \to -\infty$
present two colliding plane waves propagating along $t-x$ and $t+x$ axis (see \cite{Feinstein2}, where
the dependence between initial conditions for Veneziano's scenario and spacetime structure near the 
singularity has been determined from the classical point of view).

Let us complicate the problem a bit: imagine that initial conditions are described by a bath of plain waves
propagating along a fixed surface. It turns out that this quasi-three-dimensional problem also can be solved exactly
by means of inverse scattering method \cite{BZ}. The corresponding quantum problem hardly can be solved, too,
as well as the full four-dimensional problem, both classical and quantum one.

Nevertheless, since the classical and quantum dynamics of generic perturbations for Veneziano's scenario can be described
in terms of colliding plain waves, one should think that in the case of general initial conditions
the universe acquires a stochastic domain structure near the singularity, with proper anisotropic regime of expansion
being realized in each domain.

%%%%%%%%%%%%%%%%%%%%%%%%%%%%%%%%%%%%%%%%%%%%%%%%%%%%%%%%%%%%%%%%%%%%%%%%%%%%%%
\subsection*{Acknowledgments}
%%%%%%%%%%%%%%%%%%%%%%%%%%%%%%%%%%%%%%%%%%%%%%%%%%%%%%%%%%%%%%%%%%%%%%%%%%%%%%

I am grateful to A.A. Starobinsky for helpful discussions and A. Feinstein for turning my attention
to the paper \cite{Feinstein2}.
This study was supported in part by the
Russian Foundation for Basic Research (project
no. 02-02-16817 and MAC 02-02-06914) and the
Basic Research Program ``Nonstationary Phenomena in
Astronomy'' (Russian Academy of Sciences).

%%%%%%%%%%%%%%%%%%%%%%%%%%%%%%%%%%%%%%%%%%%%%%%%%%%%%%%%%%%%%%%%%%%%%%%%%%%%%%

\end{document}